\documentstyle[aps]{revtex}

\begin{document}
\draft

\title{ Transition from Poissonian to GOE
level statistics in a modified Artin's billiard}

\author{A. Csord\'as}
\address{Research Institute for Solid State Physics\\
P. O. Box 49, H-1525 Budapest, Hungary}

\author{R. Graham}
\address{Fachbereich Physik, Universit\"at-Gesamthochschule, Essen\\
P. O. Box 103764, D4300 Essen 1, Federal Republic of Germany}

\author{P. Sz\'epfalusy}
\address{Institute of Solid State Physics, E\"otv\"os University \\
M\'uzeum krt. 6-8., H-1088 Budapest, Hungary and \\
Research Institute for Solid State Physics\\
P. O. Box 49, H-1525 Budapest, Hungary}

\author{G. Vattay\cite{*}}
\address{Institute of Solid State Physics, E\"otv\"os University \\
M\'uzeum krt. 6-8., H-1088 Budapest, Hungary}

\date{\today}
\maketitle

\widetext
\begin{abstract}
One wall of Artin's billiard on the Poincar\'e
half plane is  replaced by a one-parameter ($c_p$) family
of nongeodetic walls.
A brief description of the classical phase space of this
system is given.
In the quantum domain,
the continuous and gradual
transition from the Poisson like to GOE level statistics
due to the small perturbations breaking the symmetry responsible
for the 'arithmetic chaos' at $c_p=1$ is studied.
Another GOE $\rightarrow$ Poisson transition due to the
mixed phase space for large perturbations is also investigated.
A satisfactory description of the intermediate level statistics
by the Brody distribution was found in both cases. The study
supports the existence of a scaling region around $c_p=1$.
A finite size scaling relation
for the Brody-parameter as a function of $1-c_p$ and the number
of levels considered can be established.
\end{abstract}
%\twocolumn[
%]

\narrowtext

\section{Introduction}
\label{sec:intro}

One of the fundamental questions of quantum
chaos is the relationship between the different types
of classical motion and the energy
level statistics of the quantum counterpart of the system\cite{BG,BE,H}.
One important field for testing ideas
on the subject is the geodesic flow on constant negative
curvature surfaces \cite{BV,AS1}.
The most often studied cases
are the symmetric octagon and the modular surface
\cite{BV,AS2,CsGSz,BGGS}.
The level spacing statistics of these systems were not found
to follow the predictions of the random matrix theory for the
Gaussian orthogonal ensemble (GOE) \cite{BV,AS2,CsGSz}.
This is contrary to the general belief that
completely chaotic systems with time reversal symmetry
have GOE level spacing statistics \cite{Betall,H,LHSS}.
This unexpected feature was
explained recently \cite{BSS,BGGS} based on semiclassical
arguments. Unlike generic ergodic systems, there is an
exponential degeneracy of time periods\cite{ABS}
of closed classical orbits caused by nontrivial
symmetries \cite{BSS}, which is responsible for
the amplification of the quantum interference
in these systems\cite{BGGS}.

In this paper we investigate a modification of
Artin's billiard introduced in Ref.\cite{CsGSz},
where one geodetic wall is replaced with a one
parameter family of nongeodetic walls.
The advantage  is that changing
only one parameter we can study perturbations of
this  billiard.
The perturbation breaks the degeneracy of
time periods and the
transition from Poissonian to GOE level spacing
distribution can be followed.
At parameter values corresponding to large perturbations
another transition from GOE to Poissonian statistics
can be observed due to the change of the ergodic part of the
mixed phase space up to complete integrability\cite{GHSzV}.

The paper is organized as follows.
In Sec. \ref{sec:classics} we introduce the family of the
billiards, and give a brief description of the classical dynamics.
In Sec. \ref{sec:QM} we shall investigate
the manifestation of the classical dynamics in
the statistical properties of the energy levels. Finally, Sec.
\ref{sec:conc} is devoted to conclusions.

\section{Classical treatment}
\label{sec:classics}

Our billiard is given in the Poincar\'e half plane with
Gaussian curvature $K=-1$ and Riemannian scalar curvature $R=-2$.
The motion between two bounces
can be obtained from the classical Lagrangian
\begin{equation}
L=\frac{1}{2y^2}(\dot{x}^2 + \dot{y}^2)
\end{equation}
or Hamiltonian
\begin{equation}
H=\frac{1}{2}y^2(p_x^2 +p_y^2)
\end{equation}
where
\begin{equation}
p_x=\frac{\dot{x}}{y^2}, \;\;\; p_y=\frac{\dot{y}}{y^2}.
\end{equation}
The domain of the classical motion is the area bounded by the
vertical walls
\begin{equation}
x=0, \;\;\; x=1/2\label{W1}
\end{equation}
and the circle
\begin{equation}
x^2 + (y-1+R)^2 = R^2\label{W2}
\end{equation}
with radius $R$, and center $(0,1-R)$.
We use the inverse radius $c_p=1/R$
as a control-parameter.
The collisions with the walls are elastic.
The trajectories of the free motion between the
bounces are circles centered on the $x$ axis.
The footpoint ($x$) and the inverse radius ($M=1/r$)
of the trajectories
(Fig. \ref{f1}) is a good choice for the "bounce mapping" $T$.
The pair $(x,M)$ remains constant during the free motion
and it changes into a new pair $(x',M')$ at each bounce
\begin{equation}
(x,M) \rightarrow (x',M')\; =\; T(x,M).\label{BM}
\end{equation}

We have studied the classical dynamics in the parameter range
$0\le c_p \le 2$. Note that
for $c_p>2$ the bottom of the billiard is open.
In the classical treatment
we measure the time in  $(2E)^{-1/2}$ units. Then
the elapsed time agrees with the metric length of orbits.
The billiard at $c_p=1$ is the
domain of Artin's triangle billiard and is bounded by three geodetic
walls. At $c_p=0$ the billiard has a horizontal non-geodetic
straight segment, and the motion is integrable\cite{GSz,GHSzV}.

In the parameter range $1\le c_p < 2$ the motion is completely
chaotic,
the Lyapunov exponent is positive for each periodic orbit as
we will show next.
For $c_p=1$ this is a consequence of Artin's proof for the
ergodicity\cite{Ar}.
For other parameters we have used a
method common in the mathematical literature\cite{Bu,Don,Wo,SzD}
for the calculation of the Lyapunov exponent of the
trajectories. Namely,
following the time evolution of the local curvature $\kappa(t)$
of an orthogonal curve
along a trajectory and choosing $\kappa(0)>0$ the quantity
$\exp(\int_{0}^{T}\kappa(t)dt)$ measures the
stretching orthogonal to the orbit \cite{Bu}.
Consequently, the Lyapunov exponent of the orbit is given by the
average value of $\kappa$
\begin{equation}
\lambda=\lim_{T\rightarrow\infty}\frac{1}{T}\int_{0}^{T}\kappa(t)dt.
\label{Ljap}
\end{equation}
During the free motion $\kappa(t)$ is described
by the Riccati equation\cite{Don}
\begin{equation}
\dot{\kappa}(t)=-K(t)-\kappa^2(t) \label{Rici}
\end{equation}
where $K(t)$ is the local Gaussian curvature, which has
a constant value of $K(t)=K=-1$ in our case.
The change of $\kappa$ before and after the collision with a wall
having curvature $q$ in the place of the bounce, is given by
\begin{equation}
\kappa_+=\kappa_- - \frac{2q}{\cos{\varphi}}\label{Bounce}
\end{equation}
where $\varphi$ is the angle of the bounce.
If the  curvature
$q$ is negative or zero for each wall, it is easy to see from
the solution of (\ref{Rici}) that after
sufficient but finite time ($T_{max}$) $\kappa$
becomes larger than one\cite{Megj1}, and then $\kappa(t)\ge +1$
holds for $t\ge T_{max}$.
The average of $\kappa$ also becomes larger than $+1$ which means
that the system is hyperbolic.
This is the case when $c_p>1$, since the curvature of
the bottom wall is
\begin{equation}
q=\frac{1}{r_g}-\frac{1}{R}\label{q}
\end{equation}
where $r_g$ is the radius of the geodesic touching the
wall in the point where the curvature is computed (Fig. 1.).
The quantity (\ref{q}) is
negative or zero, if $1\le c_p=1/R$.
The Lyapunov exponent is $\lambda=1$, if
$c_p=1$, and $\lambda\ge 1$, if $c_p>1$.
Since these arguments hold for each orbit, the
system cannot contain stable periodic orbits.

In the remaining parameter range $0<c_p<1$ it is not obvious
for which parameter values the system  is completely chaotic, and where
the first island of stability is created.
The bottom wall is locally focusing in this parameter range\cite{Megj2}.
We expect that the highly unstable free motion suppresses the
focusing on a global scale up to a certain strength of the
local focusing.  In other words,
there exists a parameter $c_p^{*}<1$
so that for $c_p^{*}<c_p<2$ the system is completely chaotic.

To have an insight how the mixed phase space sets in we have
investigated analytically
the stability of a family of periodic orbits starting at the left
corner of the billiard and bouncing $m$ times on the vertical
walls (Fig. \ref{f2}).
This family of orbits exists in the whole
$0\le c_p \le 2$ parameter range. They are obviously unstable
for $c_p > 1$. We shall show later that for $c_p < 1$ their
stability criterions give a good insight into the structure of
the mixed phase space.
Varying the parameter,
the shape of these orbits remains unchanged.
The solution of the Riccati equation (\ref{Rici}) for $c_p < 1$
and $0<\kappa(0)<1$ between two collisions with the bottom wall reads as
\begin{equation}
\kappa_m(t)=\tanh(t-t^{(m)}_0)\label{Sol}
\end{equation}
where $t^{(m)}_0$ is a parameter to be adjusted.
According to (\ref{Bounce}) bouncing with the bottom wall, the
curvature changes as
\begin{equation}
\kappa_m(t+0)=\kappa_m(t-0)-Q.\label{Cond}
\end{equation}
For the family of the orbits considered here $Q$ takes the form
\begin{equation}
Q=\frac{2(1-c_p)}{\cos{\varphi_m}}\, ,
\end{equation}
where $1-c_p$ is the curvature
of the bottom wall in the bounce point, and $\varphi_m$ is
the collision angle of the orbit labelled by $m$.
A direct calculation leads to the equation
\begin{equation}
\cos{\varphi_m}=\frac{m/2}{\sqrt{1+m^2/4}}.
\end{equation}
Since according to (\ref{Ljap}) the Lyapunov exponent  requires
the average of $\kappa(t)$ only in
the limit when $T$ goes to infinity, one can calculate this
with the asymptotic solution of $\kappa(t)$ which
satisfies the periodicity condition
\begin{equation}
\kappa_m(t+T_m)=\kappa_m(t),\label{Peri}
\end{equation}
where $T_m$ is the period of the orbit given by
\begin{equation}
\cosh(T_m/2)=\sqrt{1+m^2/4}
\end{equation}
from simple geometric considerations.
Solving (\ref{Peri}) with the condition (\ref{Cond})
leads to the equation
\begin{equation}
\tanh(-t^{(m)}_0)=\tanh(T_m-t^{(m)}_0)-Q
\end{equation}
for $t^m_0$. This is a second order equation
for $\tanh(t_0)$. Its relevant solution is
\begin{eqnarray}
\tanh(t_0^{(m)})=
\frac{1}{m}\Bigl(&&-\sqrt{c_p^2(m^2+4)-4c_p} \nonumber \\
&&+(1-c_p)\sqrt{4+m^2}\Bigr)
.\label{tant}
\end{eqnarray}
The Lyapunov exponent is the average of $\kappa$ for one period:
\begin{eqnarray}
\lambda_m&=&\frac{1}{T_m}\int_0^{T_m}\tanh(t-t_0^{(m)})dt \nonumber\\
&=&\frac{1}{T_m}\ln(\cosh(T_m-t_0^{(m)})/\cosh(t^{(m)}_0)).\label{lam}
\end{eqnarray}
{}From Eq. (\ref{lam}) and (\ref{tant}) one can calculate the
Lyapunov exponent as a function of $c_p$ and the index of
the orbit $m$ analytically. Fig. \ref{f3} shows the
Lyapunov exponent as a function of $c_p$ for this family.
For $c_p=1$ all the orbits have the same Lyapunov exponent
($\lambda_m=+1$) as a well known consequence of the geodetic walls.
Using Eqs. (\ref{lam}) and (\ref{tant}) one can also solve the
equation
\begin{equation}
\lambda_m(c_p)=0
\end{equation}
and get that parameter value $c_p$ where the $m$-th orbit
of the family gets stabilized when $c_p$ is decreased
\begin{equation}
c_p=\frac{4}{m^2 + 4}.
\end{equation}
First the shortest orbit ($m=1$) becomes stable at
$c_p=0.8$.

To decide wether there exists another periodic orbit, which is stable for
a larger parameter value $c_p$, we have investigated the
bounce mapping (\ref{BM}) (see Fig. \ref{f4}).
We have not found any island in the
range $0.8<c_p<1$. The existence of a stable orbit
causing a very small island, of course,
cannot be excluded, but the phase space in this
range can be considered completely chaotic at least practically.

%Changing the parameter from $c_p=0.8$ to $c_p=0$ continuously,
%we can observe  monotonically increasing number of
%stable islands, and a decreasing chaotic fraction of the
%phase space. It can be noticed, that
%there exists a value $M(c_p)$ so that the region $\mid M \mid <M(c_p)$
%in the phase space of the bounce mapping (\ref{BM}) does not
%contain stable islands. We can represent the chaotic
%fraction of the phase space by the value $M(c_p)$ at a given
%parameter $c_p$ (see Fig. 5.).

After this description of the classical domain,
we turn to the quantum treatment of the
system.

\section{Level statistics and scaling properties}
\label{sec:QM}

The energy eigenvalues of the billiard can be computed
by solving the Schr\"o\-din\-ger equation
\begin{equation}
-\frac{1}{2}y^2\left(\frac{\partial^2 \psi}{\partial x^2}+\frac{\partial^2
\psi}{\partial y^2}\right)=E\psi\label{Sch}
\end{equation}
with Dirichlet ($\psi=0$) boundary condition on the walls (\ref{W1}) and
(\ref{W2}). In the integrable case ($c_p=0$) the partial differential
equation (\ref{Sch}) is separable in $x$ and $y$, and the eigenfunctions
can be derived\cite{GHSzV}. Using this functional basis
for the non-integrable problem  the Schr\"odinger equation (\ref{Sch})
can be solved numerically. For further details
we refer to our earlier papers on this model\cite{GSz,GHSzV,CsGSz}.
The unfolding of the levels has been performed by the transformation
\begin{equation}
x_i=\bar{N}(E_i)\label{unf},
\end{equation}
where $\bar{N}(E)$ is given by the Weyl formula of Ref.\cite{CsGSz}.

Once one has the unfolded eigenenergies, one can analyze their fluctuations.
We concentrated on the distribution $p(s)$ of spacings
between nearest neighbour unfolded levels ($s=x_{i+1}-x_{i}$),
since this has been found to be very
sensitive the classical dynamics\cite{McD,BG,LHSS}.
The distribution $p(s)$ and its first momentum are normalized
due to the unfolding (\ref{unf}):
\begin{equation}
\int_0^{\infty}p(s)ds=1, \;\;\; \int_0^{\infty} sp(s)ds=1.\label{norm}
\end{equation}
It is expected, that the low lying energy levels are
determined by the quantum nature of the system while
the higher semiclassical part of the spectrum reflects
the classical properties, and exhibits universal behavior for
integrable and completely chaotic systems.
Generic integrable systems show Poissonian distribution\cite{Sinai,BT,Cas,Sel}
\begin{equation}
p(s)=e^{-s}.\label{Poisson}
\end{equation}
In the completely chaotic case the spectra can be described by
ensembles of random matrices, namely the spacing distribution of
the GOE applies\cite{DyM}.
This distribution is almost identical with the Wigner surmise
\begin{equation}
p(s)=\frac{\pi s}{2}e^{-\pi s^2/4}.\label{GOE}
\end{equation}
In the case of mixed phase space the situation is less clear,
the fluctuations are non-universal, and the discussion
about relevant parameters characterizing the level spacing
distribution of these systems is not finished yet.

It is of particular interest that at $c_p=1$ the level statistics do
not follow  the generic behavior of classically chaotic systems.
In this 'arithmetic' case \cite{BGGS,BSS} there exist
arguments for the relevance  of
the Poissonian distribution
based on the Selberg trace formula and on the asymptotic
distribution of time periods of periodic orbits\cite{BGGS}.
However, these considerations are valid for the asymptotic, infinitely
high part of the spectrum. In the statistics of the first
say 2000 levels deviations from the Poissonian statistics can
be observed \cite{CsGSz,BGGS}.
For parameters slightly different from
$c_p=1$ the universal GOE distribution should show up
immediately in the asymptotic, infinitely high part
of the spectrum, if the universality of GOE for
completely chaotic systems is true. The change from
Poissonian to GOE will gradually extend towards the lower lying
levels, when further decreasing $c_p$.

According to the classical behaviour of the system and
the considerations above, we can expect
three regions in the parameter space where the nearest neighbour level
spacing distribution of a finite number of levels behaves differently.
The first  region is $0\le c_p<0.8$, where with the exception of the
integrable case $c_p=0$ the phase space
is mixed one can expect some intermediate statistics
between GOE and Poissonian.
The second region is $0.8<c_p<2,\;\; c_p\neq 1 \pm \Delta$, where GOE
is expected.
The narrow region around $c_p=1$, indicated by
$\Delta$  is a third region,
where we should observe certain deviations from the
GOE for any finite number of levels due to the 'arithmetic chaos' in
$c_p=1$.

The first and third regions require a familiy of distributions
interpolating between the Poissonian and the GOE
for quite different reasons.
There are several attempts to find a theoretically well established
transition formula for systems with mixed phase space\cite{BR,Haake}.
One of the main differences in these cross-over formulas is in the
small $s$ behaviour of $p(s)$.
The Berry-Robnik formula \cite{BR} gives a constant
value for small $s$ ($p(s)\sim const$), while the Lenz-Haake formula
\cite{Haake} leads to a linear
level repulsion $s$ ($p(s)\sim s$).

There are other formulas which
interpolate between (\ref{GOE}) and (\ref{Poisson})
on a pure empirical base. One of them was proposed by
Israilev\cite{Is} and another one by Brody\cite{Brody}.
Since they are not necessarily associated with a mixed phase space they
can be tried also in the third region.
Both of them show power law level repulsion for
small $s$ ($p(s)\sim s^{\beta}$). To characterize intermediate
distributions we have chosen the one parameter family of
Brody distributions since it is simpler than the other one and,
for other purposes, we could not see any advantage of Israilev's distribution.
Moreover, in a recent study\cite{Robnik} both of them were
found equally satisfactory. The one parameter family of Brody
distributions is  given by
\begin{equation}
p(s,\beta)=a s^{\beta}\exp(-bs^{\beta +1}) \, ,
\end{equation}
where $a$ and $b$ can be derived from the conditions (\ref{norm}) as a
function of the Brody parameter $\beta$
\begin{equation}
a=(\beta +1)b,\;\;\;
b=\left(\Gamma\left(\frac{\beta+2}{\beta+1}\right)\right)^{\beta +1}.
\end{equation}
For $\beta=0$ (\ref{Poisson}) and for $\beta=1$
(\ref{GOE}) are recovered. The cumulative level density of
the Brody distribution is
\begin{equation}
I(s)=\int_0^s p(x)dx=1-\exp(-bs^{\beta+1}).
\end{equation}
We have fitted Brody distributions to our empirical $I(s)$ functions
applying the method of least squares. Using the first $n$ energy levels at
a certain $c_p$ and finding the best fit
yields an empirical $\beta_n(c_p)$ value.
Note that we do not claim, that the Brody distribution is the
correct asymptotic form, but only that the fitted $\beta$ parameter
is a good measure  where the intermediate distribution is
lying between (\ref{Poisson}) and (\ref{GOE}).
We have carried out this program in the parameter region $0\le c_p \le 1.2$,
calculating about 2000 energy levels.
We have found that the Brody distrribution provides a satisfactory fit
in the whole parameter
region%\footnote{Andr\'as, Something about the poor fit of Berry-Robnik.}

In Fig. \ref{f6}., where the Brody parameter $\beta$ as a function of $c_p$ is
depicted, one can see that $\beta$ increases continuously
with $c_p$ in the region $0\le c_p <0.8$, reflecting the fact that
in the phase space the volume of the chaotic region grows up.
This is in complete
agreement with earlier findings of Ref.\cite{Shudo} in another
system with mixed phase space.
%can be compared to Fig. \ref{f5}..
%We can see, that the Brody parameter and
%the volume of chaotic fraction of the phase space are proportional in
%the mixed phase space region  This is in complete
%agreement of earlier findings of Ref.\cite{Shudo} in an other
%system with mixed phasespace.
The saturation of
$\beta$ starting at $c_p\approx 0.5$ indicates, that the level spacing
distribution is not sensitive to small stable islands.
A relatively sharp transition from GOE toward Poissonian
can be observed in the $0.1>\mid 1-c_p\mid$ parameter range, which
we have studied in more detail.

In the critical point $c_p=1$ the numerical level spacing distribution
is still far from the theoretical Poissonian distribution.
To see the convergence of the Brody parameter towards $\beta(c_p=1)=0$
we have plotted $\beta_n(c_p=1)$ as a function of the first $n$ energy
levels (Fig \ref{f7}.). The convergence is quite apparent and $\beta_n$
can be well approximated
in the range $n=200-2000$ as
\begin{equation}
\beta_n(c_p=1)=13.05 n^{-\alpha}\label{b1scal}.
\end{equation}
with $\alpha \approx 0.55$.
This is very close to a $\sim 1/\sqrt{n}$ dependence, which is
typical in random matrix description of transitions between
ensembles\cite{Guhr}.
When $c_p \equiv 1-\delta < 1$ we observed for sufficiently small
$\delta$ values ($\delta < 0.01$) the following behavior of $\beta$ as
function of $n$. For small $n$ numbers ($n<n^*(\delta)$) the
$\beta_n(c_p=1-\delta)$ curves coincide and they tend to constant
values for $n>n^*(\delta)$ within the $n<2000$ range (Fig. \ref{f8}).
These values are scaling with $\delta$ as
\begin{equation}
\beta_n(1-\delta)\sim \delta^{\gamma}\label{bscal}
\end{equation}
where $\gamma \approx 0.44$.
The separate scaling relations (\ref{b1scal}) and (\ref{bscal})
can be unified into one finite size scaling relation familiar from
the theory of phase transitions
\begin{equation}
\frac{1}{\beta_n(1-\delta)}=\delta^{-\gamma}f(1/\delta
n^{\epsilon}),\label{scalhip}
\end{equation}
where this relation defines the scaling function $f(x)$.
For $x\rightarrow 0$ the scaling function goes to a
finite value, while for $x\rightarrow \infty$ it behaves like
$f(x)\sim x^{-\gamma}$. The $n$ dependence (\ref{b1scal})
can be recovered by choosing $\epsilon=\alpha/\gamma$.
The existence of such a formula is a stronger evidence of
the scaling then the two scaling relations (\ref{b1scal}) and
(\ref{bscal}) together, since the scaling functions computed
from different data must coincide.
This has been checked at three different $\delta$
parameters. Satisfactory agreement with the scaling
hypothesis (\ref{scalhip}) has been found within the
statistical fluctuations of the scaling functions (see Fig. \ref{f9}).

Since our billiard system is expected to be completely chaotic for
small $\delta$ values a
finite $\beta\neq 1$ asymptotic value would mean that the
GOE statistics is not universal.
We think instead, that the finite value
of $\beta$ is restricted to a region
$n^{*}(\delta) \ll n \ll n^{**}(\delta)$ only.
The emergence of such a region can be understood in terms
of semiclassical considerations.
The strength of the perturbation of the arithmetic billiard is proportional
to $\delta=1-c_p$.  Let us denote by $S_p(E)$
the action belonging  to the periodic orbit p.
The perturbation destroys the degeneracy of
the actions $S_p(E)$.
If two actions
in the arithmetic billiard are degenerate
$S_p(E)=S_{p'}(E),\;\;p\neq p'$, they differ in an amount of
\begin{equation}
\Delta S_{p,p'}(\delta,E)=S_p(\delta,E)-S_{p'}(\delta,E)=
(2E)^{1/2} l_{p,p'}(\delta)\label{deg}
\end{equation}
in the perturbed
case, where $ l_{p,p'}(\delta)=l_p(\delta)-l_{p'}(\delta)$
is the difference of the
metric length of the periodic orbits $p$ and $p'$.
$l_{p,p'}(\delta)$ is independent of the energy.
The absence of the degeneracy of the periodic orbits $p$ and $p'$
can be observed in physical quantities when the difference
of actions is comparable to $\hbar$ (in our units $\hbar =1$):
\begin{equation}
\Delta S_{p,p'}(\delta,E)\approx \hbar.\label{degcond}
\end{equation}
Because of the energy dependence (\ref{deg}) of this relation,
at low energies (small $n$) the absence of the degeneracy of
actions cannot be observed. The condition (\ref{degcond}) gives a
critical energy value to each pair of periodic orbits, and
their minimum defines $n^*\sim E_{min}/d$,
where $d$ is the mean level spacing\cite{CsGSz}.
This argument also explains that all the
$\beta_n(1-\delta)$ curves coincide for $n\ll n^*$.
For $\delta=0$ this region extends to the
infinite $n$ range. When $\delta>0$ for sufficiently high energy (large $n$)
the broken degeneracy of all periodic orbits can be observed
and the system looks like an ordinary completely
chaotic system. This defines another scale $n^{**}$ depending on $c_p$.
Between $n^*$ and $n^{**}$ we expect to observe
roughly a constant value of $\beta_n$.
{}From $n^{**}$, the $\beta_n$ function is
supposed to converge toward $\beta=1$.
For $c_p\rightarrow 1$ $n^{**}$ will grow
to infinity, and the length of the scaling region is
increasing too. However, for $c_p$ sufficiently far
from $1$ the scaling region is very restricted
or disappears ($n^{*}\approx n^{**}$).
At parameter $\delta=0.03$ instead of a plateau,
$\beta_n$ has merely a minimum around $n\approx 700$
(Fig.\ref{f10}).

\section{Conclusions}
\label{sec:conc}

In this paper we have analyzed the connection of classical dynamics
and the statistical properties of the quantum energy spectrum
in a constant negative curvature billiard family with a nongeodetic
wall.
In cases, where the phase space is mixed, we have found remarkable
agreement with the Brody distribution which interpolates between
GOE and Poissonian. The Brody
parameter was found to be a monotonously increasing function of the
chaoticity of the system  as measured by the magnitude of the
positive Lyapunov exponent. Using the Brody parameter we
could describe a GOE $\rightarrow$ Poissonian transition around
$c_p=1$ due to the arithmetic chaos.
We observed a finite size scaling relation among the Brody parameter,
the number of energy levels and the parameter $1-c_p$,
and argued that the existence of the scaling region
can be explained by the energy dependence of the condition
(\ref{degcond}). We hope, the detailed quantitative theory
of this scaling behavior can be worked out in the future.

We are indebted to D. Sz\'asz, A. Kr\'amli and
T. Guhr%\footnote{Add other names if necessary}
for enlightening discussions.
This work were supported by the National Scientific Research
Foundation under the Grant Nos. OTKA 2090, F4472, F4286 and within the
framework of the German-Hungarian Scientific and Technological
Cooperation under Project 62 "Investigation of classical and
quantum chaos". One of the
authors (G. V.) thanks for the financial support of the Sz\'echeny
Foundation and the hospitality of the Nonlinear Group of the
Niels Bohr Institute.

%\pagebreak

%\pagebreak

\begin{figure}
\caption[]{Our billiard model on the Poincar\'e half plane.\label{f1}}
\end{figure}

\begin{figure}
\caption[]{Periodic orbits bouncing once in the left corner
and $m$ times on the vertical walls.\label{f2}}
\end{figure}

\begin{figure}
\caption[]{ Lyapunov exponents as a function of the
$c_p$ parameter, for the periodic orbits of Fig. 2..
Only nonzero Lyapunov exponents are plotted.\label{f3}}
\end{figure}

\begin{figure}
\caption[]{The phase space of the bounce mapping (\ref{BM})
for $c_p=1,\;0.5,\;0.07,\;$ are shown. Notice, that there
is no island of stability for a certain $M(c_p)>\mid M\mid$
strip.\label{f4}}
\end{figure}

%\begin{figure}
%\caption[]{The $M(c_p)$ as a function of $c_p$. $M(c_p)$ characterizes
%the chaoticity of the phase space.\label{f5}}
%\end{figure}

\begin{figure}
\caption[]{The Brody parameter as a function of the control parameter
$c_p$. Each point, denoted by a $+$ sign was calculated from
the first 1947 quantum levels.\label{f6}}
\end{figure}

\begin{figure}
\caption[]{The Brody parameter as a function of the number of levels on
a log-log plot at parameter $c_p=1$ corresponding to the
arithmetic case. The dashed line represents the best power-law fit
relation $\beta_n=13.05 n^{-0.55}$.\label{f7}}
\end{figure}

\begin{figure}
\caption[]{The Brody parameter as a function of the number of levels.
The curves are taken at $c_p=1.000,0.996,0.995$ and $0.990$.
The bigger the $c_p$ the lower lying the corresponding curve.
\label{f8}}
\end{figure}

\begin{figure}
\caption[]{The scaling function $f(x)$ reconstructed from
$\beta_n(c_p)$ at parameters $c_p=0.990,0.995$ and $0.996$.
Notice that the curves coincide within their fluctuations.\label{f9}}
\end{figure}

\begin{figure}
\caption[]{The Brody parameter as a function of the number of
levels at $c_p=0.970$. Notice the verry narow scaling region around
$n=700$.\label{f10}}
\end{figure}


\begin{references}
\bibitem{Megj1} In reality, it is enough to show that $\kappa
(0)> +1$ implies $\kappa(t) \ge +1$ for $t>0$, which is more obvious.
\bibitem{Megj2} The bottom
wall (a circle), would be dispersing in an Euclidian space,
but here it is focusing due to the metric.
\end{references}
\end{document}